\documentclass[%
aps,
prb,
twocolumn
]{revtex4-2}

\usepackage[dvipdfmx]{graphicx}
\usepackage{mathtools}
\usepackage{amsmath}
\usepackage{amssymb}

\begin{document}


\title{Real-space topological characterization of quasiperiodic quantum walks: Boundary-dependent phases and the Schur index}


\author{F. Iwase}
\email[]{iwasef@tokyo-med.ac.jp}
\affiliation{Department of Physics, Tokyo Medical University\\
 6-1-1 Shinjuku, Shinjuku-ku, Tokyo 160-8402, Japan}



\begin{abstract}
We study the topological properties of one-dimensional discrete-time quantum walks with Fibonacci quasiperiodic modulation.
Spectral analysis under open boundary conditions reveals isolated edge modes that coexist at both zero and $\pi$ energies within bulk gaps.
Using the mean chiral displacement (MCD) as a dynamical bulk probe, we obtain a fractal, butterfly-like phase diagram, indicating a nontrivial bulk topology.
However, since the MCD involves wave-packet averaging, its direct correspondence with boundary-localized states remains ambiguous.
To resolve this issue, we determine the integer topological invariant by employing the Schur function formalism, a scattering-based approach defined on the unit disk.
This analysis identifies a topological phase with winding number $W=2$ inside the ``wings'' of the butterfly diagram, demonstrating that the coexistence of zero- and $\pi$-energy edge modes is an intrinsic bulk property.
Moreover, we find that the topological index crucially depends on the phason degree of freedom: the winding number can take values of $W=0$, $2$, and even $4$ depending on the local surface termination.
This behavior originates from the surface impedance, which yields an effective winding number ranges from complete masking by reflective boundaries ($W=0$) to enhancement through surface resonances ($W=4$).
Finally, we show that although individual surface terminations lead to anisotropic phase diagrams, their ensemble average restores the overall geometric structure observed in the MCD phase diagram.
Our results establish a complete bulk--edge correspondence in quasiperiodic quantum walks and provide a guiding principle for surface topological engineering, where edge transport can be controlled purely through boundary manipulation.
\end{abstract}


\maketitle

\section{Introduction\label{sec:intro}}
Discrete-time quantum walks (DTQWs) provide a versatile framework for simulating topological phases of matter~\cite{Kitagawa2010, Kitagawa2012, Asboth2016}.
As periodically driven (Floquet) systems governed by unitary time-evolution operators, DTQWs host symmetry-protected topological phases characterized by, for example, chiral symmetry.
A defining feature of Floquet topological systems is the possibility of symmetry-protected edge states at quasienergies $E=0$ and $E=\pi$, reflecting the periodicity of the quasienergy spectrum~\cite{Kitagawa2010, Asboth2013}.
This property distinguishes Floquet topological insulators from static Hamiltonian systems and enables access to topological phenomena that have no equilibrium counterpart.

While topological phases in translationally invariant systems are well understood within band theory and by topological invariants defined in momentum space, such as winding numbers and Chern numbers~\cite{HasanKane2010,Qi2011,Chiu2016}, this framework ceases to apply in the absence of lattice periodicity.
Recent interest has therefore shifted toward disordered and quasiperiodic systems, where Bloch's theorem breaks down~\cite{Levine1984, Shechtman1984}.
Quasiperiodic systems, including those generated by the Fibonacci sequence, interpolate between crystalline order and randomness.
They are characterized by self-similar, fractal energy spectra---exemplified by Hofstadter butterflies~\cite{Hofstadter1976, Kohmoto1983}---and by critical wavefunctions that are neither fully extended nor exponentially localized~\cite{AubryAndre1980, Hiramoto1989}.

These properties pose fundamental challenges for defining and diagnosing topology beyond momentum-space formulations.
In the absence of translational symmetry, the conventional definition of topological invariants based on the Brillouin zone is no longer applicable.
To address this issue, a variety of real-space approaches have been developed, including topological markers such as the Bott index~\cite{Loring2010} and the local Chern marker~\cite{Kitaev2006, Bianco2011}, as well as dynamical probes exemplified by the mean chiral displacement (MCD)~\cite{Cardano2017, Maffei2018}.
Using these tools, it has been demonstrated that quasiperiodic systems can host nontrivial topological phases, including analogs of the integer quantum Hall effect~\cite{Kraus2012}.

However, a fundamental question remains concerning the bulk--edge correspondence in quasiperiodic systems.
Unlike random disorder, quasiperiodic lattices exhibit deterministic long-range order encoded in the phason degree of freedom---a global phase shift that uniquely determines the local lattice configuration~\cite{Socolar1986}.
Recent studies have indicated that topological edge states in such systems can be highly sensitive to the specific boundary termination~\cite{Madsen2013,Kraus2012}.
Notably, in quasiperiodic systems the boundary configuration is not an independent choice but is uniquely fixed by the value of the phason~\cite{Socolar1986}.
While dynamical probes such as MCD provide access to global bulk topology, their inherent wave-packet averaging obscures the detailed correspondence between bulk topological invariants and the microscopic boundary conditions.

In this work, we establish a rigorous bulk--edge correspondence in one-dimensional Fibonacci quantum walks by combining dynamical MCD with a scattering-based Schur function analysis~\cite{Cedzich2018, Cedzich2022}.
The Schur function formalism, defined on the unit disk, enables the computation of integer-valued topological invariants directly from a semi-infinite boundary, providing a real-space characterization of edge topology.
Our main findings are threefold: 
(i) The bulk MCD reveals a fractal topological phase diagram with a Hofstadter-butterfly-like structure, within which well-defined nontrivial topological regions emerge.
(ii) The edge topology exhibits a pronounced dependence on the local surface termination: reflective boundaries lead to a masking effect with $W=0$, whereas resonant boundaries can enhance the winding number up to $W=4$.
(iii) The phase diagram obtained by the Schur analysis, combined with an ensemble average over boundary-dependent invariants, reproduces the global phase diagram obtained from the bulk MCD.
Together, these findings demonstrate that the macroscopic bulk topology of quasiperiodic systems is holographically encoded in the ensemble of all admissible boundary terminations, establishing a guiding principle for surface topological engineering based solely on boundary control.

\section{Model and Method\label{sec:method}}
\subsection{One-Dimensional Quantum Walk}

We consider a DTQW on a one-dimensional lattice with $N$ sites~\cite{Aharonov1993}.
The Hilbert space is given by $\mathcal{H}=\mathcal{H}_P\otimes \mathcal{H}_C$, where $\mathcal{H}_P$ is spanned by the position basis $\{|x\rangle\}_{x\in\mathbb{Z}}$ and $\mathcal{H}_C$ denotes the two-dimensional coin space spanned by $\{|L\rangle$, $|R\rangle\}$, representing the internal (chirality) degrees of freedom.

The time evolution is governed by a unitary operator $U=SC$, such that a single step of the walk is defined by
\begin{equation}
    |\psi(t+1)\rangle = U |\psi(t)\rangle.
\end{equation}
The shift operator $S$ conditionally translates the walker according to its chirality:
\begin{equation}
    S = \sum_x \left( |x-1\rangle \langle x| \otimes |L\rangle \langle L| 
    + |x+1\rangle \langle x| \otimes |R\rangle \langle R|\right).
\end{equation}
The coin operator $C$ acts locally on the internal degree of freedom and is defined as
\begin{equation}
    C = \sum_x |x\rangle \langle x|\otimes R(\theta_x),
\end{equation}
where $R(\theta_x)$ is a site-dependent rotation matrix,
\begin{equation}
    R(\theta_x) = 
    \begin{pmatrix}
        \cos\theta_x & \sin\theta_x \\
        -\sin\theta_x & \cos\theta_x
    \end{pmatrix},
\end{equation}
and $\theta_x$ denotes the rotation angle at lattice site $x$.

To investigate the bulk--edge correspondence, we compute the quasienergy spectrum by exact diagonalization of the unitary time-evolution operator $U$ under open boundary conditions (OBC).
At the boundaries of the finite chain, we impose reflective boundary conditions to strictly preserve the unitarity of $U$, ensuring that all eigenvalues lie on the unit circle in the complex plane.
The quasienergy $E$ is defined through the eigenvalue equation 
\begin{equation}
    U|\psi\rangle = e^{-iE}|\psi\rangle .
\end{equation}

\subsection{Quasiperiodic Modulation}

To introduce quasiperiodicity, we modulate the coin parameter $\theta_x$ according to a Fibonacci sequence~\cite{Kohmoto1983}, generated by the substitution rules $\mathrm{A} \to \mathrm{AB}$ and $\mathrm{B} \to \mathrm{A}$, starting from the seed letter $\mathrm{A}$.
This iterative construction produces the standard Fibonacci sequence, which begins as ABAABAB$\ldots$ from the left boundary.

The coin angles are assigned according to the sequence as
\begin{equation}
    \theta_x = 
    \begin{cases}
        \theta_\mathrm{A} & \text{if the $x$-th symbol is $\mathrm{A}$}, \\
        \theta_\mathrm{B} & \text{if the $x$-th symbol is $\mathrm{B}$},
    \end{cases}
\end{equation}
where $\theta_\mathrm{A}$ and $\theta_\mathrm{B}$ are two distinct rotation angles associated with the letters $\mathrm{A}$ and $\mathrm{B}$, respectively.

Numerically, we consider a finite lattice of length $N$, typically chosen as a Fibonacci number $F_n$ to ensure that the quasiperiodic modulation is naturally accommodated within the finite system.

Within the cut-and-project description of the Fibonacci sequence, different truncation points under OBC correspond to different choices of the global phase of the projection window, commonly referred to as the phason angle.
Accordingly, varying the boundary termination amounts to sampling distinct realizations of the same bulk quasiperiodic order related by phason shifts.
In the Schur function analysis presented below, we systematically vary these surface terminations to elucidate how the edge topology responds to the phason degree of freedom.

\subsection{Dynamical Topological Marker: Mean Chiral Displacement}

The system possesses chiral symmetry, represented by an operator $\Gamma$ satisfying
\begin{equation}
    \Gamma U \Gamma^{-1} = U^\dagger .
\end{equation}
In the coin basis $\{|L\rangle, |R\rangle\}$, where the local state at site $x$ is expressed as a spinor $|\psi_x\rangle = [L_x, R_x]^T$, the chiral symmetry operator is defined as the Pauli-$\sigma_x$ matrix acting on the coin space,
\begin{equation}
    \Gamma = \bigoplus_x \sigma_x
    = \bigoplus_x 
    \begin{pmatrix}
        0 & 1 \\
        1 & 0
    \end{pmatrix}.
\end{equation}

To dynamically probe the bulk topological properties, we employ the MCD~\cite{Cardano2017, Maffei2018}.
The MCD quantifies the displacement of a wave packet resolved by chiral sectors.
For a walker initially localized at the origin, the MCD after $t$ time steps is defined as:
\begin{eqnarray}
    C_\mathrm{MCD}(t) 
    &=& 2\langle\psi(t) | \hat{x} \otimes \sigma_z | \psi(t)\rangle \\
    &=& 2\sum_x x \left(|L_x(t)|^2 - |R_x(t)|^2\right).
\end{eqnarray}
Here, $\hat{x}=\sum_x x |x\rangle \langle x|$ is the position operator acting on the spatial degree of freedom, and the factor of $2$ is introduced so that the MCD value converges to integer topological invariants.
The overall sign convention depends on the choice of coin basis; however, different conventions yield identical topological information.

In chiral-symmetric systems, $C_\mathrm{MCD}(t)$ converges in the long-time limit to a quantized value in the gapped topological phases, thereby serving as a real-space dynamical indicator of the bulk topology.
To construct the bulk phase diagram, we compute the long-time average $\overline{C_\mathrm{MCD}}$.
In numerical simulations for a finite system of size $N$, the walker is initialized at the center of the lattice, and the MCD is averaged over a time window $T$ chosen such that boundary effects are avoided ($T < N/2$).
Unless otherwise stated, we use a sufficiently large system size $N = F_{18} = 2584$, ensuring that reflections from the boundaries do not affect the measurement.

\begin{figure}
\begin{center}
\includegraphics[width=8.5cm]{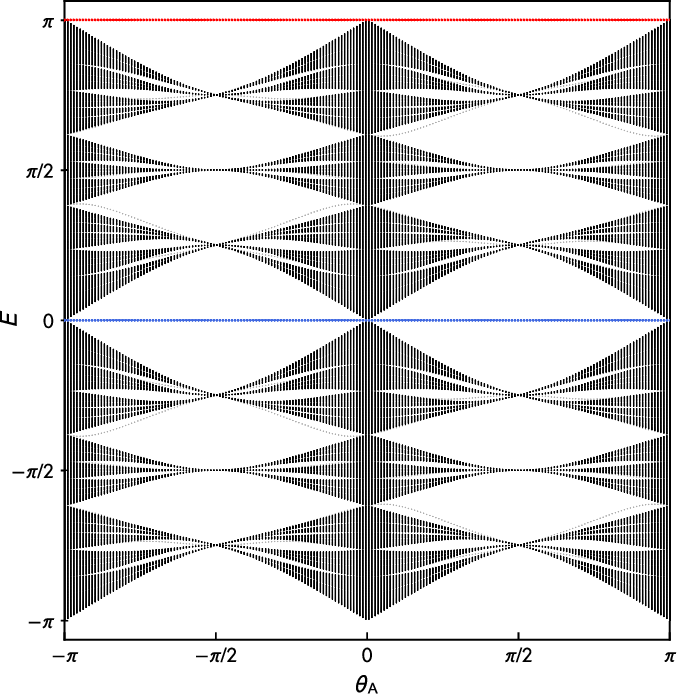}%
\caption{\label{fig:Spectrum} 
Quasienergy spectrum of the one-dimensional Fibonacci quantum walk under open boundary conditions.
The quasienergies $E$ are shown as a function of the coin parameter $\theta_\mathrm{A}$ for a finite system of size $N=F_{18}=2584$, with $\theta_\mathrm{B}$ fixed to zero.
The bulk spectrum (black dots) displays a self-similar fractal structure characteristic of quasiperiodic modulation.
Within the major quasienergy gaps, topologically protected edge states emerge: 
zero-energy modes pinned at $E=0$ (blue dots) and anomalous Floquet modes pinned at $E=\pi$ (red dots).
Quasienergies are defined modulo $2\pi$, such that states at $E=-\pi$ are physically equivalent to those at $E=+\pi$.
}
\end{center}
\end{figure}

\subsection{Schur Analysis of Boundary Scattering}

While the MCD provides a global topological signature, it inherently averages over the boundary detail and does not directly resolve individual edge modes in real space.
To achieve a rigorous characterization of topological edge states, we employ a scattering-based formulation in terms of Schur functions, following the framework developed for quantum walks by Cedzich \textit{et al.}~\cite{Cedzich2018, Cedzich2022}.
This approach establishes a direct correspondence between bulk topology and the reflection properties of the semi-infinite system.

The Schur function $f(z)$ is defined as the reflection coefficient of the semi-infinite quantum walk, regarded as a function of a complex spectral parameter $z$ inside the unit disk $|z|\leq 1$.
Physically, $f(z)$ encodes the response of the system to an incoming wave from the boundary and serves as a powerful probe of topological edge modes.

To compute $f(z)$, we introduce a sequence of auxiliary Schur functions $\{f_n(z)\}$, where $f_n(z)$ denotes the reflection coefficient of the subsystem extending from site $n$ to infinity.
These functions satisfy a recursive relation known as the Schur algorithm:
\begin{equation}
    f_n(z) = \frac{\gamma_n + z f_{n+1}(z)}{1 + \gamma_n z f_{n+1}(z)},
\end{equation}
where $\gamma_n=\cos\theta_n$ is the local reflection amplitude at site $n$, determined by the quasiperiodic modulation of the coin angle $\theta_n$ according to the Fibonacci sequence.

The recursion is initialized with the boundary condition $f_N(z)=0$ for a sufficiently large cutoff $N$, corresponding to a transparent boundary at infinity.
By iterating the Schur algorithm backward from $n=N$ to the boundary at $n = 0$, we obtain the Schur function of the full semi-infinite system, which we identify as $f(z)\equiv f_0(z)$.

In chiral-symmetric systems belonging to class AIII, the topological invariant is encoded in the winding of the boundary reflection amplitude~\cite{Ryu2010}.
In the Schur-function formulation of DTQWs, this invariant is naturally associated with the phase winding of the Schur function $f(z)$ defined on the unit circle.

In previous works, the topological index has been inferred from the value of $f(z)$ at symmetry-protected points $z=\pm 1$~\cite{Cedzich2018}.
While this prescription is sufficient for phases with $|W|\leq 1$, it may become ambiguous in the presence of higher winding numbers, where phase wrapping can obscure the distinction between, for example, $W=0$ and $W=2$.

Since the quasiperiodic Fibonacci quantum walk studied here exhibits a hierarchy of nontrivial topological phases with $|W|\geq 2$, we adopt the full winding-number definition based on the analytic structure of the Schur function,
\begin{equation}
    W = \oint_{|z|=1} \frac{dz}{2\pi i}\frac{f_0'(z)}{f_0(z)}.
\end{equation}
Hereafter, we refer to $W$ as the Schur index, or equivalently the Schur winding number.

This definition provides a rigorous and unambiguous characterization of the edge topology, capturing the global phase structure of the boundary scattering amplitude even under strong quasiperiodic modulation.
Geometrically, $W$ counts the number of times the trajectory of $f_0(z)$ winds around the origin in the complex plane as $z$ traverses the unit circle.

\section{Results\label{sec:results}}

\begin{figure}
\begin{center}
\includegraphics[width=8.5cm]{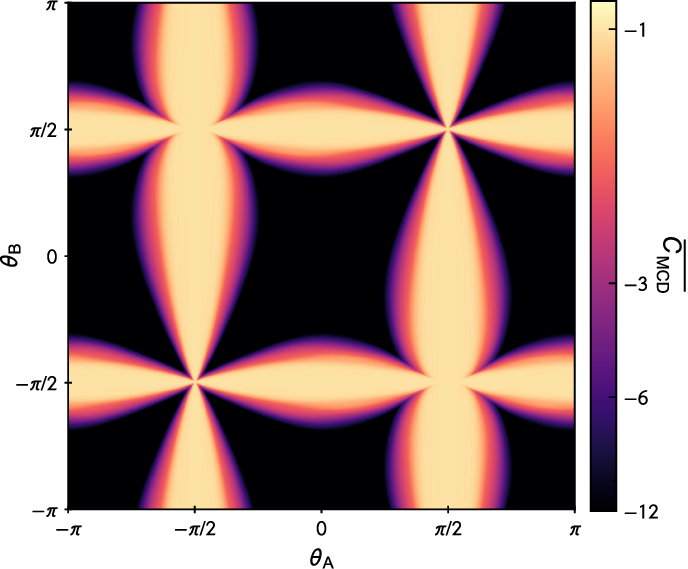}%
\caption{\label{fig:MCD_PD} 
Dynamical bulk topological phase diagram obtained from the long-time averaged mean chiral displacement (MCD).
The MCD is mapped in the $(\theta_\mathrm{A}, \theta_\mathrm{B})$ parameter plane.
Bright regions (yellow) correspond to gapped topological phases, where the MCD converges to quantized value ($\overline{C_\mathrm{MCD}} \approx -1$), signaling a nontrivial bulk topology.
Dark regions (black to purple) indicate metallic or critical regimes, in which the wavepacket undergoes anomalous spreading and the MCD grows to large negative values (down to $\sim -1200$).
For clarity, the color scale is saturated at $-12$. 
The resulting phase diagram exhibits a self-similar ``butterfly-like'' fractal structure, reflecting the quasiperiodic nature of the underlying Fibonacci modulation and mirroring the structure of the quasienergy spectrum.
}
\end{center}
\end{figure}

\subsection{Bulk Properties and Edge Modes in Closed Systems}

We begin by examining the bulk spectral properties of the Fibonacci quantum walk in a closed geometry.
Figure~\ref{fig:Spectrum} displays the quasienergy spectrum of a finite system with size $N=F_{18}=2584$ under OBC as a function of the coin parameter $\theta_\mathrm{A}$, with $\theta_\mathrm{B}$ fixed to zero.
The spectrum exhibits a pronounced self-similar fractal structure, reflecting the underlying quasiperiodic order of the Fibonacci modulation.

Within the quasienergy gaps, two distinct classes of isolated modes are clearly resolved: zero-energy modes at $E=0$ and anomalous Floquet modes pinned at $E=\pm\pi$.
These states are localized near the system boundaries and are indicative of nontrivial topological phases protected by chiral symmetry.
Their coexistence highlights a hallmark feature of Floquet topological systems, absent in static Hamiltonian counterparts.

To characterize the bulk topology independently of boundary-specific details, we calculate the long-time average of the MCD, $\overline{C_\mathrm{MCD}}$, as a dynamical bulk topological marker.
The resulting phase diagram, shown in Fig.~\ref{fig:MCD_PD}, reveals a hierarchy of quantized plateaus (bright regions, $\overline{C_\mathrm{MCD}} \approx -1$) associated with large quasienergy gaps,
Remarkably, these plateaus organize into a butterfly-like fractal structure reminiscent of the Hofstadter spectrum, centered at $(\theta_\mathrm{A}, \theta_\mathrm{B})=(\pm \pi/2, \pm \pi/2)$.
This structure identifies extended parameter regions where the bulk resides in a nontrivial topological phase.

While the MCD phase diagram provides a clear dynamical fingerprint of bulk topology, it intrinsically averages over the spatial extent of the wavepacket.
As a consequence, it does not uniquely resolve the integer-valued topological invariant associated with individual edge modes observed in the quasienergy spectrum.
This limitation motivates a complementary boundary-sensitive analysis, which we address in the following subsection.

\subsection{Surface-Dependent Edge Topology}

\begin{figure}
\begin{center}
\includegraphics[width=8.5cm]{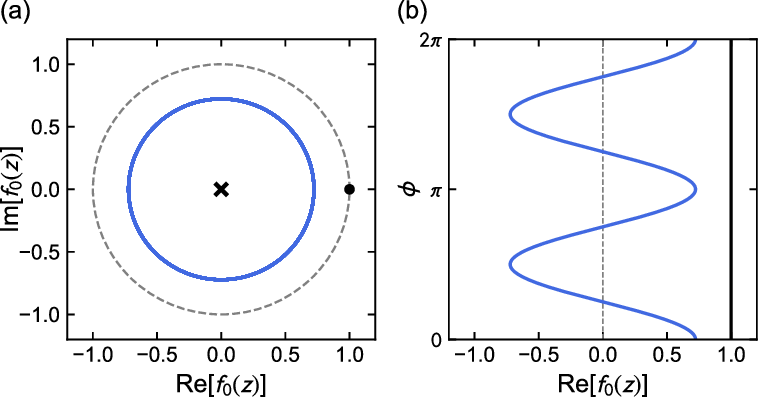}%
\caption{\label{fig:trajectory_schur}
Topological characterization of boundary modes via Schur functions.
(a) Trajectories of the Schur function $f_0(z)$, interpreted as the macroscopic reflection coefficient of the semi-infinite system, calculated at the parameter point $(\theta_\mathrm{A}, \theta_\mathrm{B})=(\pi/2, 0)$ as $z$ traverses the unit circle $|z|=1$.
For a reflectionless surface termination (Type~A, site-A termination, blue), the trajectory forms a perfect circular loop centered at the origin and winds around it twice, reflecting a robust nontrivial boundary topology deep inside the quasienergy gap.
In contrast, for a perfectly reflective termination (Type~B, site-B termination, black), the trajectory is strictly pinned at the point $(1,0)$, corresponding to unit reflection and illustrating the surface masking effect, whereby boundary topology becomes trivial despite a nontrivial bulk.
(b) Real part of the Schur function, $\mathrm{Re}[f_0(z=e^{i\phi})]$, plotted as a function of the phase $\phi$ (``unwrapped'' representation).
For Type-A termination, the curve exhibits two full oscillations over $\phi\in[0, 2\pi)$, crossing zero four times and yielding a winding number $W=2$.
For Type-B termination, $\mathrm{Re}[f_0(z)]$ remains strictly positive and featureless, indicating the absence of winding ($W=0$).
}
\end{center}
\end{figure}

While the MCD provides a global dynamical signature of bulk topology, it does not spatially resolve individual edge states nor uniquely determine the boundary topological invariant.
To explicitly characterize the edge topology in real space, we therefore employ the Schur function analysis introduced in Sec.~\ref{sec:method}.

We focus on a representative parameter point $(\theta_\mathrm{A}, \theta_\mathrm{B})=(\pi/2, 0)$, which lies deep inside a nontrivial topological phase identified by the MCD plateau (Fig.~\ref{fig:MCD_PD}).
This choice allows us to demonstrate the bulk--edge correspondence without ambiguities arising from finite-size gap closing or critical scattering effects.
Throughout this analysis, we consider a finite system of size $N=F_{13}=233$, which is sufficient to faithfully capture the boundary scattering properties.

\begin{figure}
\begin{center}
\includegraphics[width=8.5cm]{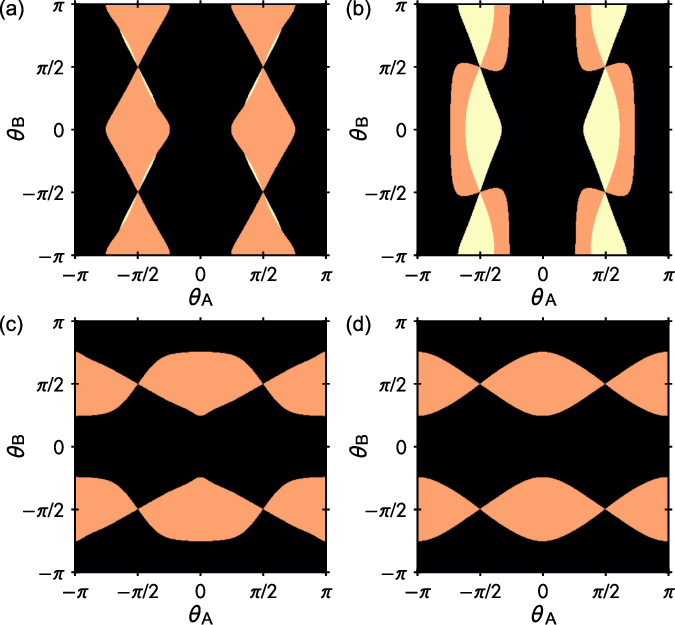}%
\caption{\label{fig:Schur_PDs}
Surface-dependent local topological phase diagrams obtained from the Schur index.
Maps of the Schur winding number $W$ in the $(\theta_\mathrm{A}, \theta_\mathrm{B})$ parameter plane for distinct surface terminations:
(a) ABA, (b) AAB, (c) BAA, and  (d) BAB.
The color coding is as follows: black denotes the trivial phase ($W=0$), orange indicates the topological phase with $W=2$, and yellow highlights regions with an enhanced winding number $W=4$.
(a) For ABA termination, vertically oriented diamond-shaped regions with $W=2$ appear along the lines $\theta_\mathrm{A}= \pm \pi/2$.
Narrow regions with a higher winding number $W=4$ emerge near the boundaries of these domains.
(b) The AAB termination exhibits similar vertical structures centered around $\theta_\mathrm{A}= \pm \pi/2$, but with noticeably distorted geometries.
In this case, the $W=4$ phase is substantially broadened compared to panel (a), reflecting a pronounced surface resonance effects.
(c), (d) In contrast, the BAA and BAB terminations display horizontally oriented lobes centered around $\theta_\mathrm{B}= \pm \pi/2$.
These phase diagrams are dominated by the $W=2$ phase and are characterized by smoother, more rounded boundaries.
}
\end{center}
\end{figure}

Figure~\ref{fig:trajectory_schur}(a) shows the trajectory of the Schur function $f_0(z)$ in the complex plane as the complex parameter $z=e^{i\phi}$ traverses the unit circle.
For the site-A termination (labeled as Type~A, blue), where the boundary is effectively transparent due to $\gamma_\mathrm{A}=0$ at $(\theta_\mathrm{A}, \theta_\mathrm{B}) = (\pi/2, 0)$, the trajectory forms a perfect circle centered at the origin and encloses it.
This winding behavior is further illustrated in Fig.~\ref{fig:trajectory_schur}(b), which plots the real part $\mathrm{Re}[f_0(z)]$ as a function of the phase $\phi$.
The Type-A termination exhibits two full oscillations over $\phi\in[0, 2\pi)$, corresponding to a winding number $W=2$.

This result demonstrates that the reflectionless boundary faithfully reveals the intrinsic bulk topology predicted by the MCD.
In stark contrast, for a termination obtained by replacing the leftmost letter of the standard Fibonacci sequence from A to B (Type~B, $\gamma_\mathrm{B}=1$), the Schur trajectory remains pinned near $\mathrm{Re}[f_0(z)]=1$ and fails to encircle the origin, yielding a trivial winding number $W=0$.
This behavior exemplifies a pronounced masking effect: a perfectly reflective boundary effectively decouples the edge modes from the scattering probe, rendering the observed boundary topology trivial despite the nontrivial topology of the bulk.

We evaluate the Schur winding number $W$ across the $(\theta_\mathrm{A}, \theta_\mathrm{B})$ parameter plane for four distinct surface terminations: ``ABA$\ldots$'', ``AAB$\ldots$'', ``BAA$\ldots$'', and ``BAB$\ldots$''.
These terminations are constructed by modifying only the first three letters of the standard Fibonacci sequence, while preserving the quasiperiodic bulk structure.
Importantly, this choice is exhaustive at the level of three-site environments: the above sequences constitute the complete set of locally allowed boundary configurations in the Fibonacci lattice.

As shown in Figs.~\ref{fig:Schur_PDs}(a)--(d), the resulting boundary-resolved phase diagrams differ strikingly from one another.
Across the entire parameter space, the Schur winding number takes discrete values $W=0$, $2$, and $4$, corresponding respectively to a trivial phase, a bulk-induced topological phase, and a surface-enhanced topological phase.

A pronounced anisotropy emerges depending on the identity of the leftmost site.
For surfaces starting with letter `A' [Figs.~\ref{fig:Schur_PDs}(a) and~\ref{fig:Schur_PDs}(b)], the phase boundaries are predominantly controlled by $\theta_\mathrm{A}$, giving rise to vertically elongated structures centered around $\theta_\mathrm{A} = \pm \pi/2$.
In contrast, surfaces terminated by `B' [Figs.~\ref{fig:Schur_PDs}(c) and~\ref{fig:Schur_PDs}(d)] exhibit horizontally oriented patterns governed by $\theta_\mathrm{B}$, with dominant features aligned near $\theta_\mathrm{B} = \pm \pi/2$.
This anisotropy directly reflects the local scattering properties imposed by the surface termination.

Most remarkably, for the ``ABA$\ldots$'' and ``AAB$\ldots$'' terminations [Figs.~\ref{fig:Schur_PDs}(a) and~\ref{fig:Schur_PDs}(b)], we identify parameter regions characterized by an enhanced winding number $W=4$.
For the ABA termination, the $W=4$ phase appears only in narrow regions adjacent to the boundaries of the $W=2$ domains, whereas for the AAB termination the $W=4$ regions are substantially broadened, indicating a stronger surface-induced effect.

This enhancement originates from a surface resonance mechanism: the ``AA'' doublet at the boundary forms a Fabry--P\'erot--like resonator.
Although an individual `A' site is highly reflective, the resonant double layer allows the wave function to penetrate more deeply into the bulk.
As a result, the boundary contributes an additional phase winding of $+2$, which adds to the intrinsic bulk winding number $+2$, yielding a total Schur index of $W=4$.
By contrast, no such enhancement is observed for B-terminated surfaces, whose phase diagrams are dominated by the $W=2$ phase.

\subsection{Restoration of Bulk--Edge Correspondence}

\begin{figure}
\begin{center}
\includegraphics[width=8.5cm]{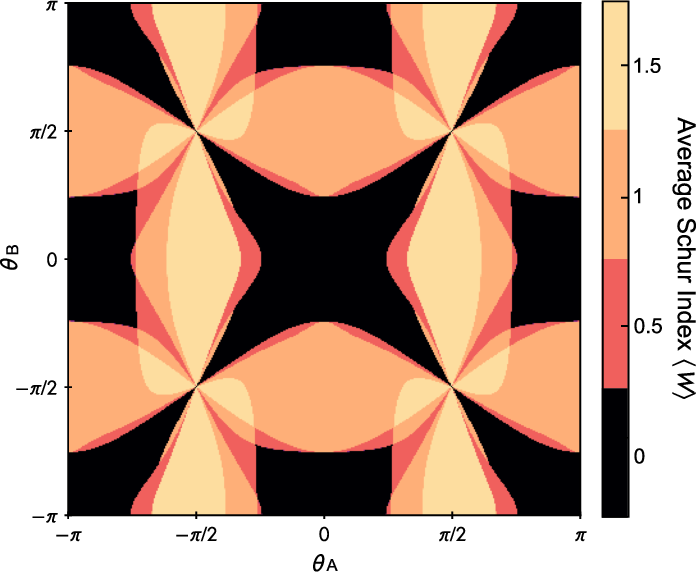}%
\caption{\label{fig:average_PD} 
Restoration of bulk--edge correspondence by ensemble averaging.
Phase diagram of the ensemble-averaged Schur winding number, $\langle W \rangle$, computed by averaging boundary winding number over all inequivalent surface terminations (phasons configurations) of the same bulk quasiperiodic lattice.
The resulting map qualitatively captures the characteristic fractal ``butterfly'' structure of the bulk dynamical phase diagram extracted from the mean chiral displacement shown in Fig.~\ref{fig:MCD_PD}.
In contrast to the strictly integer-valued winding numbers observed for individual surface terminations, the averaged index $\langle W \rangle$ develops robust plateaus at characteristic fractional values (e.g., $\langle W \rangle = 1.5$, $1$, $0.5$, and $0$).
This fractionalization reflects the statistical nature of the bulk topology in quasiperiodic systems and demonstrates that the macroscopic bulk invariant emerges as the ensemble expectation value of surface-dependent boundary invariants.
}
\end{center}
\end{figure}

The strong dependence of the Schur winding number $W$ on the surface termination naturally raises a fundamental question: what quantity should be regarded as the ``true'' bulk topological invariant in a quasiperiodic system?

To address this issue, we compute in Fig.~\ref{fig:average_PD} the ensemble-averaged Schur winding number $\langle W \rangle$, obtained by averaging $W$ over all possible phason configurations, i.e., over inequivalent surface cuts of the same bulk quasiperiodic lattice.
The resulting averaged Schur map closely resembles the fractal ``butterfly'' structure of the bulk dynamical phase diagram extracted from the MCD (Fig.~\ref{fig:MCD_PD}).
The agreement extends to the fine geometric details of the fractal patterns across the parameter space.
A slight asymmetry between the two phase diagrams is visible, which can be attributed to the dynamical nature of the MCD observable.
Nevertheless, their overall structure and phase boundaries are in excellent agreement.

Importantly, while the Schur winding numbers for a fixed surface termination are strictly quantized to integer values, the ensemble-averaged quantity $\langle W \rangle$ generically assumes robust fractional values.
These fractional plateaus reflect the statistical distribution of distinct boundary topological sectors associated with different phason configurations, rather than a breakdown of topological quantization.
In this sense, $\langle W \rangle$ should be interpreted as the expectation value of the boundary invariant over the ensemble of admissible surface realizations of the same bulk quasiperiodic structure.

This striking correspondence demonstrates that, although individual boundary realizations may exhibit strongly surface-dependent topology, the macroscopic bulk topology is recovered as the statistical expectation value of the boundary invariants.
Our results therefore establish a generalized bulk--edge correspondence for quasiperiodic systems, in which the bulk topological invariant emerges only after averaging over the ensemble of boundary terminations associated with the phason degree of freedom.

\section{Discussion\label{sec:discussion}}
\subsection{Physical Mechanism of Boundary Sensitivity}

Our results reveal an unusually strong sensitivity of topological edge modes to the local surface termination in quasiperiodic systems.
Although the bulk topology is unambiguously detected by the MCD, the explicit manifestation of edge topology, quantified by the Schur winding number $W$, depends critically on the microscopic boundary configuration.

We interpret the ``masking effect'' at reflective boundaries ($W=0$) in terms of an impedance mismatch at the surface.
When the boundary termination effectively forms a hard wall (e.g., $\gamma \approx 1$), the quasiperiodic bulk is strongly decoupled from the vacuum.
As a result, the topological edge modes fail to hybridize with the scattering channel, and the reflection coefficient remains topologically trivial despite the nontrivial bulk.
In contrast, for sufficiently transparent boundaries ($\gamma < 1$), the internal wave functions can leak into the external scattering channel.
This finite coupling allows the bulk topology to be encoded in the phase winding of the reflection amplitude, yielding ($W=2$).

This mechanism highlights a fundamental distinction between quasiperiodic systems and translationally invariant topological insulators.
In periodic systems, bulk--edge correspondence guarantees the appearance of protected edge states at any boundary that respects the protecting symmetry.
In quasiperiodic systems, however, the deterministic but nonperiodic long-range order encoded in the phason degree of freedom renders the boundary itself an essential part of the topological characterization.
Consequently, the existence of a nontrivial bulk invariant does not ensure the visibility of edge modes for every individual boundary realization, but rather guarantees their existence within the ensemble of all possible surface terminations.

\subsection{Origin of High Winding Numbers and Gap Labeling}

A central outcome of our work is the emergence of higher winding numbers $|W|\geq 2$ within a scalar Schur-function framework.
In conventional one-dimensional periodic systems with a finite unit cell, chiral-symmetric quantum walks typically support at most a single winding ($W=1$), reflecting the limited phase accumulation over a single Brillouin zone.
The observation of larger winding numbers therefore points to a fundamentally different topological mechanism in quasiperiodic systems.

In a Fibonacci quasicrystal, the notion of a finite unit cell breaks down, and the system is instead characterized by an effectively infinite unit cell, or equivalently by a hierarchy of high-order periodic approximants.
This structural complexity is captured by the gap labeling theorem, which states that the integrated density of states (IDS) within spectral gaps takes the form  $\mathrm{IDS}=\{p+q\tau ^{-1}\}$, where $\tau$ is the golden ratio and $p, q\in \mathbb{Z}$ label the gaps~\cite{Kohmoto1983}.
These gap labels encode the topological character of the quasiperiodic spectrum and provide a natural explanation for the integer quantization of the Schur winding number observed in our results.

The dominant $W=2$ phase corresponds to major spectral gaps associated with low-order gap labels.
Physically, this reflects the fact that a wavepacket traversing the self-similar quasiperiodic potential accumulates a geometric phase that winds multiple times around the origin in the complex reflection plane.
In contrast to periodic systems, where phase winding is bounded by the Brillouin-zone topology, the hierarchical structure of the Fibonacci lattice allows for repeated phase accumulation across multiple length scales, thereby enabling higher winding numbers.

The appearance of an enhanced $W=4$ phase for the ABA and AAB terminations further highlights the interplay between bulk topology and boundary physics.
In these cases, the local ``AA'' doublet at the surface acts as a Fabry--P\'erot--like resonator.
Since the `A' site acts as a reflectionless layer ($\gamma_\mathrm{A}<1$) and `B' as a reflective wall ($\gamma_\mathrm{B}\approx 1$), the consecutive `AA' sequence forms an extended surface cavity.
The incident wave function deeply penetrates this cavity and undergoes multiple reflections before escaping.
This resonance effectively samples two consecutive topological interfaces, doubling the accumulated phase winding from $W=2$ to $W=4$.
Therefore, the associated resonance induces an additional phase winding that is purely surface-generated and adds constructively to the intrinsic bulk winding.
Importantly, this enhancement occurs without modifying the bulk Hamiltonian, demonstrating that surface engineering in quasiperiodic systems can locally amplify the observable topological index while leaving the bulk invariant unchanged.

\subsection{Restoring Bulk--Edge Correspondence}

The pronounced dependence of the Schur winding number on the surface cut directly reflects the phason degree of freedom intrinsic to quasiperiodic systems.
Different boundary terminations correspond to distinct realizations of the same bulk quasiperiodic order, related by phason shifts that do not alter the bulk Hamiltonian but strongly affect local boundary properties.

Our results show that, although individual surface terminations yield Schur phase diagrams that lack the global symmetry of the bulk, their ensemble average over all inequivalent phason configurations faithfully reconstructs the bulk dynamical phase diagram obtained from the MCD.
This observation leads to a generalized formulation of bulk--edge correspondence in quasiperiodic systems: the macroscopic bulk topology is not encoded in any single boundary realization, but rather in the statistical expectation value of boundary topological invariants.

From this perspective, the ensemble averaging procedure effectively samples the phason degree of freedom, restoring the global topological structure that is obscured at the level of individual surface cuts.
The bulk invariant, therefore, emerges only after averaging over the full family of boundaries associated with the same quasiperiodic bulk.
This statistical bulk--edge correspondence provides a natural resolution to the apparent conflict between robust bulk topology and strongly surface-dependent boundary signatures in quasiperiodic lattices.

From an experimental perspective, performing an ensemble average over boundary configurations may at first appear impractical, as it seemingly requires fabricating multiple samples with distinct terminations.
However, this concern is mitigated by the principle of isomorphism of the Fibonacci lattice: the number of inequivalent local boundary environments is strictly finite.
For instance, at the level of three-site terminations considered here, only four distinct patterns exist.
Consequently, the ensemble average relevant for restoring the bulk invariant does not require sampling an extensive statistical distribution.
Instead, averaging over a small and strategically chosen set of fundamental boundary configurations---such as the four representative terminations analyzed in this work---is already sufficient to achieve rapid convergence toward the bulk topological invariant.
This finiteness renders the statistical bulk--edge correspondence not merely a conceptual construction, but an experimentally accessible procedure.
This statistical viewpoint also has direct experimental implications.

\subsection{Experimental Implications}

Our results suggest a practical route for topological surface engineering in platforms like photonic waveguides~\cite{Cardano2017} and ultracold atoms~\cite{Weitenberg2011, Atala2013}.
By utilizing local optical control to effectively redefine the termination site---conceptually equivalent to phason shift---the topological winding number can be switched (e.g., from $W=0$ to $W=2$ or $W=4$) without modifying the bulk Hamiltonian.
This boundary-driven tunability, distinct from conventional insulators, paves the way for dynamically programmable devices that harness the strong localization of edge modes and the flexible controllability of surface terminations.

\section{Conclusion\label{sec:conclusion}}

In this work, we have investigated the topological phases of Fibonacci quasiperiodic quantum walks, with particular emphasis on the interplay between the dynamical bulk signatures and local boundary geometry.
By combining the MCD as a global probe of bulk topology with a Schur function analysis as a local probe of edge physics, we have formulated a generalized bulk--edge correspondence applicable to quasiperiodic systems.

Our study leads to three central conclusions.
First, the bulk MCD exhibits a fractal, butterfly-like phase diagram, demonstrating the existence of nontrivial topological phases protected by chiral symmetry in the absence of translational invariance.
Second, the Schur function formalism reveals that the visibility and character of topological edge modes depend sensitively on the surface termination.
In particular, we identified a masking effect, whereby reflective boundaries suppress the edge winding ($W=0$), as well as a resonant amplification mechanism, in which specific terminations enhance the winding number to higher integers (up to $W=4$).
This behavior shows that in quasiperiodic media the topological invariant determined locally at the edge is not a rigid consequence of the bulk Hamiltonian, but an emergent quantity controlled by the phason degree of freedom.
Third, we demonstrated that the ensemble average over all boundary terminations restores a global correspondence: the averaged winding number reproduces the overall structure of the bulk MCD phase diagram.

These results point to a paradigm shift in the understanding of topological phases in quasiperiodic systems: the bulk topology is not encoded in a single boundary realization, but is instead holographically captured by the statistical ensemble of all possible boundary terminations associated with the phason degree of freedom.
In this sense, the bulk invariant emerges only at the ensemble level, rather than being directly accessible from any individual surface.

While we focused on the Fibonacci model, our Schur function approach is general and can be applied to other quasiperiodic systems, such as the unitary almost-Mathieu operator~\cite{Cedzich2023}, which has recently attracted significant attention in experimental contexts~\cite{Lin2025}.

From a practical perspective, our findings open a new route toward topological surface engineering.
By exploiting the sensitivity of edge topology to the boundary cut, robust edge transport channels can be selectively created, suppressed, or amplified without modifying the bulk Hamiltonian.
Such boundary-based control provides a flexible platform for designing reconfigurable topological photonic and quantum devices.

\begin{acknowledgments}
The author thanks C. Cedzich for fruitful discussions and specifically for suggesting the application of the Schur analysis, which significantly improved this work.
\end{acknowledgments}


\begin{thebibliography}{99}%
\bibitem{Kitagawa2010} T. Kitagawa, M. S. Rudner, E. Berg, and E. Demler, Exploring topological phases with quantum walks, Phys. Rev. A \textbf{82}, 033429 (2010).
\bibitem{Kitagawa2012} T. Kitagawa, Topological phenomena in quantum walks: elementary introduction to the physics of topological phases, Quantum Inf. Process. \textbf{11}, 1107 (2012).
\bibitem{Asboth2016} J. K. Asb\'oth, L. Oroszl\'any, and A. P\'alyi, \textit{A Short Course on Topological Insulators} (Springer, 2016). 
\bibitem{Asboth2013} J. K. Asb\'oth and H. Obuse, Bulk-boundary correspondence for chiral symmetric quantum walks, Phys. Rev. B \textbf{88}, 121406(R) (2013).

\bibitem{HasanKane2010} M. Z. Hasan and C. L. Kane, Colloquium: Topological insulators, Rev. Mod. Phys. \textbf{82}, 3045 (2010).
\bibitem{Qi2011} X.-L. Qi and S.-C. Zhang, Topological insulators and superconductors, Rev. Mod. Phys. \textbf{83}, 1057 (2011).
\bibitem{Chiu2016} C.-K. Chiu, J. C. Y. Teo, A. P. Schnyder, and S. Ryu, Classification of topological quantum matter with symmetries, Rev. Mod. Phys. \textbf{88}, 035005 (2016).
\bibitem{Levine1984} D. Levine and P. J. Steinhardt, Quasicrystals: A New Class of Ordered Structures, Phys. Rev. Lett. \textbf{53}, 2477 (1984).
\bibitem{Shechtman1984} D. Shechtman, I. Blech, D. Gratias, and J. W. Cahn, Metallic Phase with Long-Range Orientational Order and No Translational Symmetry, Phys. Rev. Lett. \textbf{53}, 1951 (1984).
\bibitem{Hofstadter1976} D. R. Hofstadter, Energy levels and wave functions of Bloch electrons in rational and irrational magnetic fields, Phys. Rev. B \textbf{14}, 2239 (1976).
\bibitem{Kohmoto1983} M. Kohmoto, L. P. Kadanoff, and C. Tang, Localization Problem in One Dimension: Mapping and Escape, Phys. Rev. Lett. \textbf{50}, 1870 (1983).
\bibitem{AubryAndre1980} S. Aubry and G. André, Analyticity breaking and Anderson localization in incommensurate lattices, Ann. Isr. Phys. Soc. \textbf{3}, 133 (1980).
\bibitem{Hiramoto1989} H. Hiramoto and M. Kohmoto, Scaling analysis of quasiperiodic systems: Generalized Harper model, Phys. Rev. B \textbf{40}, 8225 (1989).
\bibitem{Loring2010} 
T. A. Loring and M. B. Hastings, Disordered topological insulators via C$^*$-algebras, Europhys. Lett. \textbf{92}, 67004 (2010).
\bibitem{Kitaev2006} 
A. Kitaev, Anyons in an exactly solved model and beyond, Ann. Phys. \textbf{321}, 2 (2006).
\bibitem{Bianco2011}
R. Bianco and R. Resta, Mapping topological order in coordinate space, Phys. Rev. B \textbf{84}, 241106(R) (2011).
\bibitem{Cardano2017} F. Cardano, A. D'Errico, A. Dauphin, M. Maffei, B. Piccirillo, C. de Lisio, G. Salerno, M. Lewenstein, P. Massignan, and E. Santamato, Detection of Zak phases and topological invariants in a chiral quantum walk of twisted photons, Nat. Commun. \textbf{8}, 15516 (2017).
\bibitem{Maffei2018} M. Maffei, A. Dauphin, F. Cardano, M. Lewenstein, and P. Massignan, Topological characterization of chiral models through their long time dynamics, New J. Phys. \textbf{20}, 013023 (2018).
\bibitem{Kraus2012} Y. E. Kraus, Y. Lahini, Z. Ringel, M. Verbin, and O. Zilberberg, Topological States and Adiabatic Pumping in Quasicrystals, Phys. Rev. Lett. \textbf{109}, 106402 (2012).
\bibitem{Socolar1986} 
J. E. S. Socolar and P. J. Steinhardt, Quasicrystals. II. Unit-cell configurations, Phys. Rev. B \textbf{34}, 617 (1986). 
\bibitem{Madsen2013} K. A. Madsen, E. J. Bergholtz, and P. W. Brouwer, Topological equivalence of crystal and quasicrystal band structures, Phys. Rev. B \textbf{88}, 125118 (2013).
\bibitem{Cedzich2018}
C. Cedzich, T. Geib, F. A. Gr\"{u}nbaum, C. Stahl, L. Vel\'{a}zquez, A. H. Werner, and R. F. Werner, The topological classification of One-Dimensional Symmetric Quantum Walks, Ann. Henri Poincar\'{e} \textbf{19}, 325 (2018).
\bibitem{Cedzich2022}
C. Cedzich, T. Geib, F. A. Gr\"{u}nbaum, L. Vel\'{a}zquez, A. H. Werner, and R. F. Werner, Quantum Walks: Schur Functions Meet Symmetry Protected Topological Phases, Commun. Math. Phys. \textbf{389}, 31 (2022).
\bibitem{Aharonov1993} Y. Aharonov, L. Davidovich, and N. Zagury, Quantum random walks, Phys. Rev. A \textbf{48}, 1687 (1993).
\bibitem{Ryu2010} S. Ryu, A. P. Schnyder, A. Furusaki, and A. W. W. Ludwig, Topological insulators and superconductors: tenfold way and dimensional hierarchy, New J. Phys. \textbf{12}, 065010 (2010).

\bibitem{Weitenberg2011} C. Weitenberg, M. Endres, J. F. Sherson, M. Cheneau, P. Schau\ss, T. Fukuhara, I. Bloch, and S. Kuhr, Single-spin addressing in an atomic Mott insulator, Nature \textbf{471}, 319 (2011).
\bibitem{Atala2013} M. Atala, M. Aidelsburger, J. T. Barreiro, D. Abanin, T. Kitagawa, E. Demler, and I. Bloch, Direct measurement of the Zak phase in topological Bloch bands, Nat. Phys. \textbf{9}, 795 (2013).
\bibitem{Cedzich2023} C. Cedzich, J. Fillman, and D. C. Ong, Almost everything about the unitary almost Mathieu operator, Commun. Math. Phys. \textbf{403}, 745 (2023).
\bibitem{Lin2025} Q. Lin, C. Cedzich, Q. Zhou, and P. Xue, Observation of Metal-Insulator and Spectral Phase Transitions in Aubry-Andr\'e-Harper Models, arXiv:2508.08255 (2025).






\end{thebibliography}

\end{document}